\documentclass[a4paper]{article}
\usepackage{thesis-packages}

\usepackage{multirow}

\title{Effects of non-uniform number of actions by Hawkes process \\on spatial cooperation}
\author{Daiki Miyagawa${}^{1*}$ and Genki Ichinose${}^{2}$
\ \\
\ \\
${}^{1}$
Graduate School of Science and Technology, Shizuoka University, \\3-5-1 Johoku, Naka-ku, Hamamatsu, 432-8561, Japan\\
${}^{2}$
Department of Mathematical and Systems Engineering, Shizuoka University, \\3-5-1 Johoku, Naka-ku, Hamamatsu, 432-8561, Japan\\
$^*$ Corresponding author (miyagawa.daiki.18@shizuoka.ac.jp)}

\begin{document}
\maketitle

\section*{Abstract}
The emergence of cooperative behavior, despite natural selection favoring rational self-interest, presents a significant evolutionary puzzle.
Evolutionary game theory elucidates why cooperative behavior can be advantageous for survival.
However, the impact of non-uniformity in the frequency of actions, particularly when actions are altered in the short term, has received little scholarly attention.
To demonstrate the relationship between the non-uniformity in the frequency of actions and the evolution of cooperation, we conducted multi-agent simulations of evolutionary games.
In our model, each agent performs actions in a chain-reaction, resulting in a non-uniform distribution of the number of actions.
To achieve a variety of non-uniform action frequency, we introduced two types of chain-reaction rules: one where an agent's actions trigger subsequent actions, and another where an agent's actions depend on the actions of others.
Our results revealed that cooperation evolves more effectively in scenarios with even slight non-uniformity in action frequency compared to completely uniform cases.
In addition, scenarios where agents' actions are primarily triggered by their own previous actions more effectively support cooperation, whereas those triggered by others' actions are less effective.
This implies that a few highly active individuals contribute positively to cooperation, while the tendency to follow others' actions can hinder it.


\section*{Keywords}
Evolutionary game, 
Donation game, 
Spatial structure, 
Cascading action, 
Non-uniformity

\section{Introduction}

Cooperation is one of the key elements of human society. 
However, competition among individuals frequently arises due to the potential benefits of exploiting others. 
Game theory posits that cooperation is inherently irrational, as individuals are often tempted by opportunities for exploitation.  
Thus, the reasons why cooperation is common in society have attracted great attention and many researchers continue to tackle this problem.

Evolutionary game theory is frequently used to analyze the dilemma of cooperation. 
This theoretical framework, which draws on concepts from biological evolution, examines which behaviors and traits are likely to persist across numerous generations~\cite{Maynardsmith1982book}. 
In standard evolutionary games, a population is composed of many individuals.
Each individual possesses a strategy to survive in a population and utilizes this strategy when playing a game with other individuals. 
Individuals who achieve higher payoffs than others in such interactions can reproduce more successfully, resulting in a greater number of offspring. 
Consequently, the strategies of these individuals become more prevalent within the population over time.

Many researchers have used evolutionary games to seek effective strategies for maintaining cooperation. 
For example, when considering long-term relationships, it has been theoretically demonstrated that cooperative strategies can survive~\cite{Nowak1993Nature,Fudenberg1994AmEconRev,Sigmund2010book,imhof2007tit}. 
Evolutionary game models have also revealed that being evaluated by others contributes to the evolution of cooperation~\cite{Nowak1998nature,Nowak1998jtb,OhtsukiIwasa2004jtb}. 
Additionally, the introduction of spatial structures such as lattices, small-world networks, and scale-free networks is known to facilitate the formation of clusters of cooperators, enhancing their chances of survival~\cite{Nowak1992Nature_spatial,Santos2005PRL,Ohtsuki2006Nature}.


It is well studied in evolutionary theory that the relative rate at which strategies are updated compared to the frequency of interactions can significantly influence the outcomes of evolution~\cite{roca2006time,roca2009evolutionary,wu2009diversity,liu2018influence}.
In classical evolutionary games the rate of strategy update is assumed to be much slower than the interaction.
This scenario assumes biological evolution, wherein the basic premise is that individuals alter their traits and behaviors across generations. 
Consequently, while interactions occur numerous times throughout an individual's lifetime, it is reasonable to assume that strategy updates happen at most a few times.
In cultural evolution, on the other hand, individuals modify their behavior multiple times throughout their lifetime. 
Therefore, the rate of strategy updates closely matches the frequency of interactions.
Additionally, several studies have also investigated the effects of dynamically changing the rate of strategy updates~\cite{rong2010emergence,rong2013coevolution,cong2014time,mao2021effect}.

Timescales are not the only distinguishing factor between cultural and biological evolution. 
Another key difference is the non-uniformity in the number of interactions experienced by individuals.
In biological evolution, the rate of interactions is significantly faster compared to the rate of strategy updating, resulting in a high frequency of interactions for all individuals. 
Consequently, there is relatively little variation in the number of interactions among individuals. 
However, in cultural evolution, strategy updating occurs at stages with insufficient interactions. 
As a result, the number of interactions among individuals tends to be more non-uniform than in biological evolution.

Roca {\it et al.}~considered the effect of such non-uniformity of the number of interactions among individuals on the evolution of cooperation~\cite{roca2006time}.
Their model realized the slower rate of interactions by repeating $n$ times to randomly pick up two agents and let them play games ($n$ is sufficiently smaller than the total number of agents $N$).
Hence, as $n$ increases (decreases), the distribution of the number of games approaches the uniform (non-uniform) state.
They showed that cooperation was hard to evolve in the prisoner's dilemma game as $n$ became smaller.
Additionally, Li {\it et al.}~has examined the impact of both ``spatial structure" and ``non-uniformity of interactions" on the evolution of cooperation ~\cite{Li2020NatComm}. 
They investigated how the relative rate of topological changes in a temporal network affects the evolution of cooperation in comparison to the rate of strategy updates. 
In this temporal network, topological changes activate (or deactivate) each link, meaning that only certain individuals can interact at any given time. 
Consequently, when topological changes are slow, some individuals can continue to interact until strategy updates occur, whereas rapid changes enable nearly all individuals to interact.
As a result, they demonstrated that when the rate of topological change is equivalent to the rate of strategy updates, cooperation is promoted more effectively than in cases where the rate of topological change is very fast.
In other words, they have shown that cooperation is better maintained in a non-uniform situation where topological changes are slow, allowing only some individuals to interact.


On the other hand, studies on the effects of non-uniformity on the evolution of cooperation are still limited.
This is because it is not clear whether, for example, cooperation is promoted in the case of extreme non-uniformity.
For instance, extreme non-uniformity in the number of interactions among individuals (which has been extensively studied as a bursting phenomenon~\cite{Barabasi2005Nature,VazquezA2006PhysRevE-burst,VazquezA2007PhysRevLett,Karsai2012SciRep,reynolds2011origin,wang2015burst}), could emerge due to self-exciting behavioral rules.
Therefore, understanding the relationship between the diversity of non-uniformity and the evolution of cooperation is essential for elucidating how the rules governing the timing of our actions influence the evolution of cooperation.

The aim of this study is to reproduce the diversity of non-uniformity in the number of interactions and to show its relationship with the evolution of cooperation. 
To achieve specific types of non-uniformity, we define cascading agents that act in a chain-reaction manner and conduct multi-agent simulations of evolutionary games. 
We prepare two types of cascading rules: 
(1) endogenous exciting cascades, where an action once taken excites subsequent actions of the same individual, and (2) exogenous exciting cascades, where the actions of others excite the next action of the individual.
From the perspective of achieving non-uniformity, we anticipate that the rule (1) will result in a state where a few individuals have a very high number of actions, and the rule (2) will produce groups where individuals excite each other, leading to a collective with a high number of actions. 
For instance, in the context of SNS, the rule (1) resembles a situation where posting an opinion makes one more likely to post related content again, while the rule (2) is similar to seeing friends post on social media, prompting one to post as well. 
Cascading actions thus can be considered realistic, so verifying the connection between these natural tendencies and forming a cooperative society is also a significant motivation for this study.
We generate these chain-reaction timings using the Hawkes process, a method commonly employed in studies using self-exciting behavioral models~\cite{ogata1988statistical,rambaldi2015modeling,hardiman2013critical,embrechts2011multivariate}.

\section{Overview of Hawkes processes}
Hawkes process is a part of the point process that represents occurrences of events in a time series~\cite{hawkes1971spectra,bacry2015hawkes}. 
The point process models the frequency of events using an intensity function $\lambda(t)$. 
We can interpret the function as an event occurrence rate at the time $t$. 
The probability that an event occurs once from $t$ until $t+\Delta t$ is hence $\lambda(t)\Delta t$. 

In the case that $\lambda(t)$ is a constant value $\rho\ (>0)$, we call this the stationary Poisson process. 
It means that the possibility of an event occurring is constant, independent of time and other factors. 
This well-known model is the basis of the queue theory~\cite{adan2002queueing}.
We call it just the Poisson process in the rest of this paper. 

Hawkes process is a subset of point process models, whose probability of generating a new event depends on the past events. 
Defining the time series of event occurrence in order $t_0, t_1,\cdots,t_{\ell}$, the Hawkes process can be explained by the following: 
\begin{align}
\lambda(t) = \rho + \sum_{t_{\ell}<t} g(t-t_{\ell}). 
\label{eq:gen_Hawkes}
\end{align}

\begin{figure}[hbtp]
\centering
\includegraphics[width=\columnwidth]{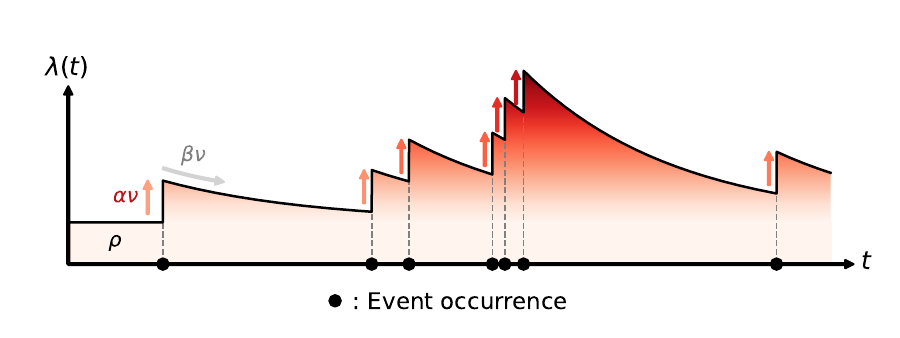}
	\caption{An example where $\lambda(t)$ fluctuates over time according to event occurrences. We use the following setting: $\rho=0.5$, $\alpha=0.5$, $\nu=1$, and $\beta=1$. }
\label{fig:hawkes_intro}
\end{figure}

\begin{figure}[hbtp]
\includegraphics[width=\columnwidth]{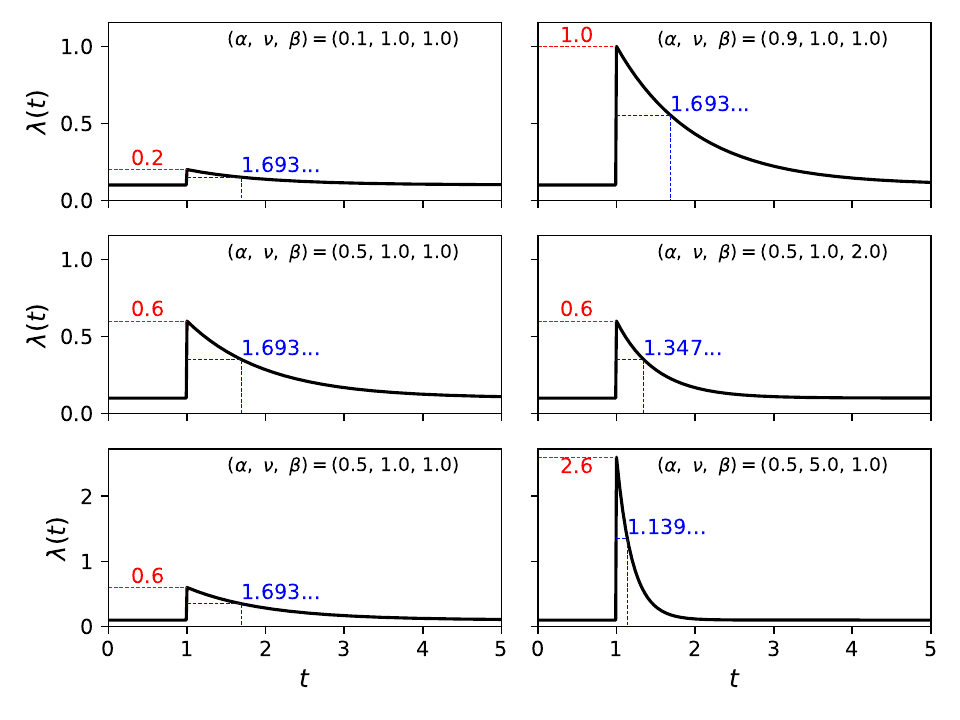}
\caption{
	Effects of the parameters on the exponential kernel (Eq.~\eqref{eq:kernel}). 
	Plots in the top row compare the effect of $\alpha$ (left: $\alpha=0.1$; right: $\alpha=0.9$). $\nu=1$ and $\beta=1$ are constant in both. 
	Plots in the middle compare the effect of $\beta$ (left: $\beta=1$; right: $\beta=2$). $\alpha=0.5$ and $\nu=1$ are constant in both. 
	Plots in the bottom compare the effect of $\nu$ (left: $\nu=1$; right: $\nu=5$). $\alpha=0.5$ and $\beta=1$ are constant in both. 
	Each red broken line denotes a peak of $\lambda(t)$ that is raised by an event occurrence. 
	Each blue broken line represents the half-life period of $\lambda(t)$ after it rises, affected by an occurrence of an event.
}
\label{fig:alpha_nu}
\end{figure}

\noindent
The function $g(\tau)$ indicates the influence that an event after $\tau$ elapsed since its occurrence has on the occurrence of a new event, 
which is called the kernel function of the Hawkes processes. 
It defines 0 if $\tau<0$ but otherwise non-negative so that exponential and power functions are often used. 
We now use the extended exponential one: 
\begin{align}
g(\tau) = \alpha \nu \exp(-\beta\nu\tau),
\label{eq:kernel}
\end{align}
where $\alpha$, $\nu$, and $\beta$ are constant parameters. 
Equation~\eqref{eq:gen_Hawkes} can be rewritten using that function as below:
\begin{align}
\lambda (t) = \rho + \sum_{t_{\ell}<t} \alpha \nu \exp\{-\beta \nu (t-t_{\ell})\}. 
\label{eq:ExpHawkes}
\end{align}
Figure~\ref{fig:hawkes_intro} shows the fluctuation of $\lambda(t)$ as an example. 
The black circles on the horizontal axis indicate the moments when events occurred ($t_{\ell}$). 
We can see that $\lambda(t)$ increases sharply at these moments by the effect of the second term in Eq.~\eqref{eq:ExpHawkes}.
Thus, the next event will more frequently occur in a few moments because $\lambda(t)$ becomes higher after an event. 
It means that the cascading of events can happen by the kernel function. 
Let us explain the parameters by this figure.
Parameter $\rho$ means the initial and the lowest value of $\lambda(t)$. 
As soon as the events occur, $\lambda(t)$ rises by the value of $\alpha\nu$ (red arrow) because of the exponential kernel. 
It then decays with steepness represented by $\beta\nu$ (gray arrow) until the next event occurs.
In other words, the larger $\alpha$ is, the larger $\lambda(t)$ rises and increases the probability of occurrence of the next event (as shown in the top row of Fig.~\ref{fig:alpha_nu}), 
while the larger $\beta$ is, the faster $\lambda(t)$ decays and the more difficult the chain of events occurs (as shown in the middle row of Fig.~\ref{fig:alpha_nu}).
Additionally, while $\nu$ is not typically included in the ordinary exponential kernel, as depicted in the bottom row of Fig.~\ref{fig:alpha_nu}, it serves as a parameter that allows for altering the spike shape of the intensity function. 
Note that changing $\nu$ does not change the expected number of events per unit time $\left<\lambda\right>=\frac{\rho}{1-\alpha/\beta}$ (which is derived in Appendix~\ref{app:expected_intensity}).
Therefore, we introduced it in our model to control the intensity of the $\lambda(t)$ change while maintaining the value of $\left<\lambda\right>$.
%

While the Hawkes process described above is one in which one's past events excite one's future events, there are also situations in which one's future events are excited by other event generators. 
These can be regarded, for example, as a situation in which one user's activity triggers another's activity in a social networking service. 
In this case, let $N_i$ be the set of neighboring event generators that can influence generator $i$, and $t_{\ell;i}$ be the $\ell$th event occurrence time of generator $i$. 
We can express the intensity function by adding up influences from all the neighbors as follows:
\begin{align}
	\lambda_i(t) = \rho_i + \sum_{j\in N_i} \sum_{t_{\ell;j}<t} \alpha_{ji} \nu_{ji} \exp\{-\beta_{ji} \nu_{ji} (t-t_{\ell;j})\}, 
\end{align}
where $\rho_i$ is generator $i$'s lowest value of $\lambda_i(t)$. 
Additionally, $\alpha_{ji}$, $\nu_{ji}$, and $\beta_{ji}$ are the parameters of the effect of the event generated by $j$ on $i$. 

\section{Model}
We developed agent-based evolutionary game models on the square lattice. 
We prepared two kinds of models. 
The first is the ``standard model", which assumes biological evolution. 
Another one is the Hawkes process model.
This model uses the Hawkes process to calculate the number of donations for each agent, but most parts overlap the standard one. 
Therefore, we first explain the standard model.

\subsection{Standard Model}
In this model, we prepare $L\times L=N$ agents separately allocated on $L\times L$ square lattice, which satisfies the periodic boundary condition. 
Each agent can interact with $k$ neighbor agents. 
Agent $i$ has its strategy $\bm s_i$. 
We define strategy $\bm s_i$ as the tuple composed of some variables that are used to make decisions. 
Basically, we use $s_i=(a_i)$, where $a_i$ indicates the attitude of agent $i$: Cooperative (C) or Defective (D); thus $a_i\in\{{\rm C}, {\rm D}\}$. 

\begin{figure*}[!b]
\centering
\includegraphics[width=\columnwidth]{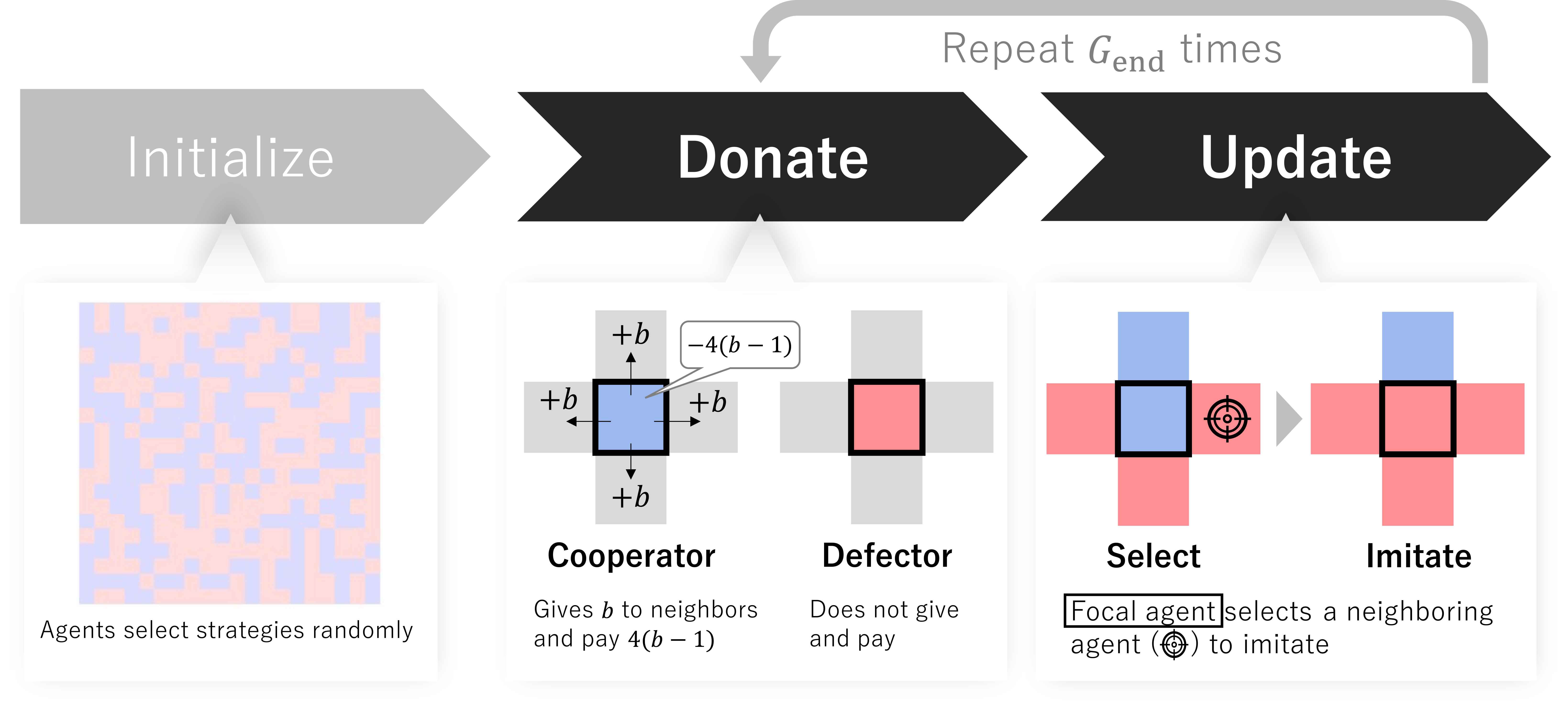}
	\caption{Flow of our model.}
\label{fig:flow}
\end{figure*}

Three stages compose the flow of our model as shown in Fig.~\ref{fig:flow}: Initialization, Donation, and Update. 
In the initialization stage, we allocate random strategies to every agent and set accumulated rewards to zero. 
In the donation stage, every agent simultaneously decides whether to donate to its neighbor agents once according to its action:
Agent $i$ with $a_i={\rm C}$, {\it i.e.}, cooperator, donates payoff $b$ to all neighbors and pays the cost $k(b-1)$. 
On the other hand, agent $i$ with $a_i={\rm D}$, {\it i.e.}, defector, donates nothing and pays nothing. 
We impose $b > 1$ to regard neighboring individuals as playing a donation game, a part of the prisoner's dilemma games, showing the payoff matrix in this game in Table~\ref{tab:payoff_matrix}.
\begin{table}[b]
	\centering
	\caption{Payoff matrix. It shows payoffs on the ``Self" player's side when the two players decide whether to cooperate.}
	\begin{tabular}{cccc}
	                      			&               	& \multicolumn{2}{c}{Opponent}		\\ \cline{3-4} 
		                      		&               	& Cooperate    & Defect   		\\ \hline
		\multirow{2}{*}{Self} 	& Cooperate        	& $1$      & $1-b$              	 	\\
	                      			& Defect 	& $b$         & $0$                		\\ \hline
	\end{tabular}
	\label{tab:payoff_matrix} 
\end{table}
In this case, parameter b is regarded as the defectors' advantage because the cost-benefit ratio becomes smaller as b becomes larger.
Each agent can store payoffs until one generation ends. 
%
In the next stage, update, every agent updates their strategies simultaneously. 
Update means to imitate a strategy of an agent whose payoff is the highest among its neighbors including itself (if there are a few candidates, the agent selects one randomly). 
When an agent imitates another agent's strategy, the agent can copy with probability $1-\mu$, but its strategy mutates with probability $\mu$ (reset each variable randomly). 
We call the consecutive process of donation and update stages the process of one generation. 
The simulation executes $G_{\rm end}$ times in one trial to obtain the final state of the individuals. 
To analyze the result, we mainly use the fraction of cooperators ($f_{\rm C}$) given by dividing the number of cooperators into that of all individuals. 
Every indicator is averaged in the last $G_{\rm ave}$ generations. 

\subsection{Hawkes process (HP) model}
The difference between the standard model and the HP one is in the donation stage. 
At each donation stage, the HP model requires agents to generate the time series of executing donations following the Hawkes process until the one-generation-time $t_G$ has elapsed.
Thus, each agent's number of actions is no longer limited to one.

The intensity function $\lambda_i(t)$ of agent $i$ is written as below:
\begin{align}
\lambda_i(t) = \rho_i &+ \alpha_{i;n} \nu_i \sum_{t_{\ell;i}<t} \exp\{-\beta_{i;n} \nu_i (t-t_{\ell;i})\} \nonumber \\
					  &+ \alpha_{i;x} \nu_i \sum_{j\in N_i} \sum_{t_{\ell;j}<t} \exp\{-\beta_{i;x} \nu_i (t-t_{\ell;j})\}, 
\label{eq:lambda_origin}
\end{align}
where $\rho_i$ is the baseline intensity, 
$\alpha_{i;n}$ and $\alpha_{i;x}$ correspond to the strength of excitation $\alpha$ in Eq.~\eqref{eq:ExpHawkes}, but they have different roles; 
The former is endogenous one, {\it {\it i.e.}} used to deal with the past events of itself, but the latter is exogenous one, {\it {\it i.e.}} used to deal with that of neighbors. 
Similarly, $\beta_{i;n}$ and $\beta_{i;x}$ correspond to the decay ratio of endogenous and exogenous, respectively. 
To simplify, we use the common parameter $\beta$ instead of all $\beta_{i;n}$ and $\beta_{i;x}$. 
We also use a common parameter $\nu$ for $\nu_i$ similarly; unless otherwise specified, we assume $\nu=1$ by default.
Variable $t_{\ell;k}$ indicates the time when agent $k$ executed its $\ell$th action.

\subsection{Algorithm to generate time series of executing donations according to Hawkes processes}

We use Ogata's thinning method to simulate generating a time series of events (which are interpreted as donation actions in our model) according to the Hawkes process~\cite{Ogata1981TransInfoTheor}. 
We show the process of generating a new event through the thinning method with a single event generator (agent) expressed by Eq.~\eqref{eq:gen_Hawkes}: 
\begin{enumerate}
\item Set $t=0$, $\lambda^*=\rho$. 
\item Regarding the process as the Poisson process with $\lambda=\lambda^*$, calculate the estimated time $t^*$ that the next event will occur using the exponential distribution. 
\item Calculate $\lambda(t+t^*)$ by Eq.~\eqref{eq:gen_Hawkes}. 
\item Execute acceptance if $u<\frac{\lambda(t+t^*)}{\lambda(t)}$; otherwise, execute rejection, where $u\in[0,1)$ is a random number following the uniform distribution. 
\begin{itemize}
	\item Acceptance: Generate and record the new event, and then update variables: $\lambda^*\leftarrow\lambda(t+t^*)+g(0)$, $t\leftarrow t+t^*$. 
	\item Rejection: Update variables without generating the new event: $\lambda^*\leftarrow\lambda(t+t^*)$, $t\leftarrow t+t^*$. 
\end{itemize}
\item Return to 2 if $t<t_G$ otherwise stop generating and go to the update stage. 
\end{enumerate}
Our algorithm is based on that thinning method and the Gillespie algorithm, which simulates the multi-variable Hawkes process regarding each agent as one event generator. 
We used the heap structure to accelerate the Gillespie algorithm. 
This improvement was proposed by the manuscript~\cite{Kiss2017EoN} and introduced into a Python library EoN, which simulated infectious disease transmission 
and was applied to the multi-variable Hawkes process by Farajtabar {\it et al.}~\cite{Farajtabar2017JMLR}. 

\section{Results}

\subsection{Experiment I: Comparing the impact of different types of cascading donation on evolution of cooperation}
First of all, we compared the HP model with the standard one to identify the effect of cascading actions on the evolution of cooperation. 
Here, the HP model is divided into three cases by the settings of $\alpha_{i;n}$ and $\alpha_{i;x}$. 
\begin{itemize}
\item Poisson-case: $\alpha_{i;n}=\alpha_{i;x}=0$ for arbitrary agent $i$; 
\item Endogenous case (Endo case): $\alpha_{i;n}=\alpha$, $\alpha_{i;x}=0$; 
\item Exogenous case (Exo case): $\alpha_{i;n}=0$, $\alpha_{i;x}=\alpha$; 
\end{itemize}
where $\alpha\in[0,1)$ is a common parameter. 
Additionally, we set $\rho$ to all agents as a common parameter instead of $\rho_i$. 
In this case, the expected value $\left<\lambda\right>$ of the intensity of occurrence for every individual for one unit of time is also common and can be obtained (see Appendix~\ref{app:expected_intensity} for details).  
We set $\left<\lambda\right>$ to one in all experiments to focus on the difference in individual donation numbers.
We can use $\rho$ to adjust $\left<\lambda\right>$.
The value $\rho$ is automatically calculated depending on $\alpha$ and $\beta$ by Eq.~\eqref{eq:adjust_rho}. 
We summarize the values of $\{\rho_i, \alpha_{i;n}, \alpha_{i;x}, \beta\}$ as shown in Table~\ref{tab:R1_params}. 
\begin{table}[btp]
\centering
	\caption{Values of parameter $\rho_i$, $\alpha_{i;n}$, $\alpha_{i;x}$, and $\beta$ in each situation. Note that the standard model doesn't use any of these parameters. }
\label{tab:R1_params}
\begin{tabular}{ccccc}
\hline
               & standard & Poisson & Endo      & Exo       \\ \hline
$\rho_i$       & -        & 1         & $1-\alpha$       & $1-\alpha$     \\ \hline
$\alpha_{i;n}$ & -        & 0         & $\alpha$         & 0              \\ \hline
$\alpha_{i;x}$ & -        & 0         & 0                & $\alpha$	   \\ \hline
$\beta$        & -        & 1         & 1                & $k$            \\ \hline
\end{tabular}
\end{table}
The excitation intensity $\nu_i$ is also unified to this experiment's common parameter $\nu$. 
In addition, we use $t_G = 1$. 
The setting combining $\left<\lambda\right> = 1$ sets each agent's expected number of actions during one donation stage to 1. 
It is intended to make the timescale of interactions close to that of strategy changes, in other words, to simulate the cultural evolution.
We used $L=100$, $G_{\rm end}=3000$, and $G_{\rm ave}=500$ to set constant parameters. 

\begin{figure}[htbp]
\centering
\includegraphics[width=0.7\columnwidth]{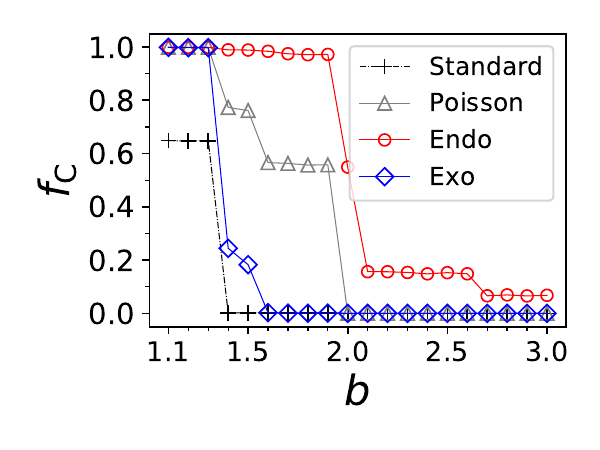}
\caption{Changes in the final average fraction of cooperators ($f_{\rm C}$) for each case. The horizontal axis represents the advantage of defectors ($b$), and the line colors distinguish between different models/cases: gray dashed line for the standard model, gray for Poisson, red for Endo, and blue for Exo cases.}
\label{fig:Result1}
\end{figure}

We investigated which case promotes cooperation, observing how cooperators evolve in each case (Fig.~\ref{fig:Result1}). 
The horizontal and vertical axes show the advantage of defectors and the final fraction of cooperators, respectively. 
The black dotted line with cross symbols corresponds to the standard model; 
the gray line with triangles corresponds to the Poisson case; 
the red line with circles corresponds to the Endo case; 
the blue line with squares corresponds to the Exo case. 
We generated each data point on the lines by averaging 100 runs.

Let us compare the standard model with the others first. 
The standard model maintains cooperators at around 65\% until $b=1.3$, beyond which it drops to zero. 
On the other hand, the others achieve a higher fraction of cooperators than the standard does. 
Poisson case maintains cooperators at around 100\% until $b=1.3$, then 75\% until $b=1.5$, 55\% until $b=1.9$, and finally, the line drops to zero.
The red line, Endo case, falls to 55\% at $b=2.0$, 15\% at $b=2.1$, and 8\% at $b=2.7$.
The blue Exo case significantly falls to 25\% at $b=1.4$ and zero at $b=1.6$. 
Hence, it shows that cooperators can survive more easily in non-uniform cases. 
Besides, the highest fraction of cooperators is achieved in the Endo case compared to others; 
the subsequent order is the Poisson and then Exo cases. 
It is implied that it is advantageous for cooperators to play the games stimulated by themselves rather than by others. 
This result showed us that acting by exciting oneself is more beneficial to maintaining a cooperative society than being encouraged by others.

\begin{figure}[htbp]
\centering
\includegraphics[width=0.7\columnwidth]{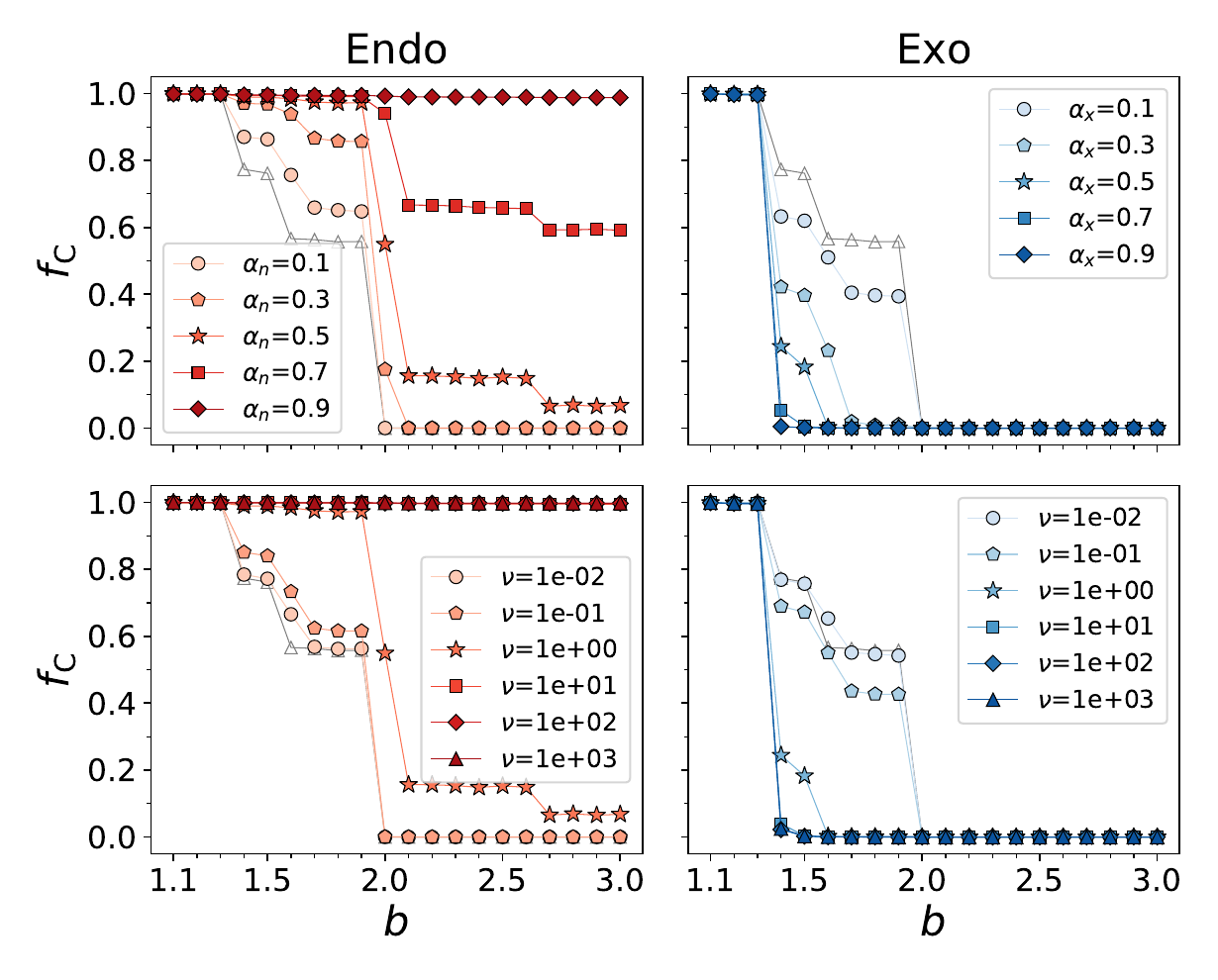}
\caption{Changes in $f_{\rm C}$ with various values of $\alpha$ and $\nu$. The upper row represents the variation with $\alpha_n$ ($\alpha_x$), while the lower row shows the variation with $\nu$. The left column (red series) corresponds to the Endo case, and the right column (blue series) corresponds to the Exo case.}
\label{fig:b-fc-various}
\end{figure}

To study the robustness of the result, we examined the dynamics of strategies in the intensity of various endogenous and exogenous cascades by changing $\alpha$ and $\nu$. 
Figure~\ref{fig:b-fc-various} clearly shows the effects of the intensity of cascades. 
The left column represents the result of the Endo case, focusing on $\alpha_n$ (upper) and $\nu$ (lower) variations while keeping $\alpha_x$ fixed at zero. 
By contrast, the right one represents the result of the Exo case. 
The axes of the four subplots are the same as those of Fig.~\ref{fig:Result1}. 
In each subplot, the outcome of the Poisson case is illustrated using gray lines and triangles. 
Let us explain these subplots one by one. 
The upper left one, which corresponds to the Endo case where we changed $\alpha$, displays that more cooperators survive as $\alpha$ increases. 
In addition, all lines are over that of the Poisson case. 
The lower left one, which corresponds to the Endo case where we changed $\nu$, provides similar trends as the upper one. 
The upper right one, which corresponds to the Exo case where we changed $\alpha$, shows that fewer cooperators survive as the parameter increases, and all lines are under that of the Poisson case. 
The lower right one, which corresponds to the Exo case where we changed $\nu$, provides similar trends as the upper one. 
These results showed that the order of the Endo, Poisson, and Exo cases is robust regardless of the intensity of cascades.

\begin{figure}[btp]
	\centering
	\includegraphics[width=\columnwidth]{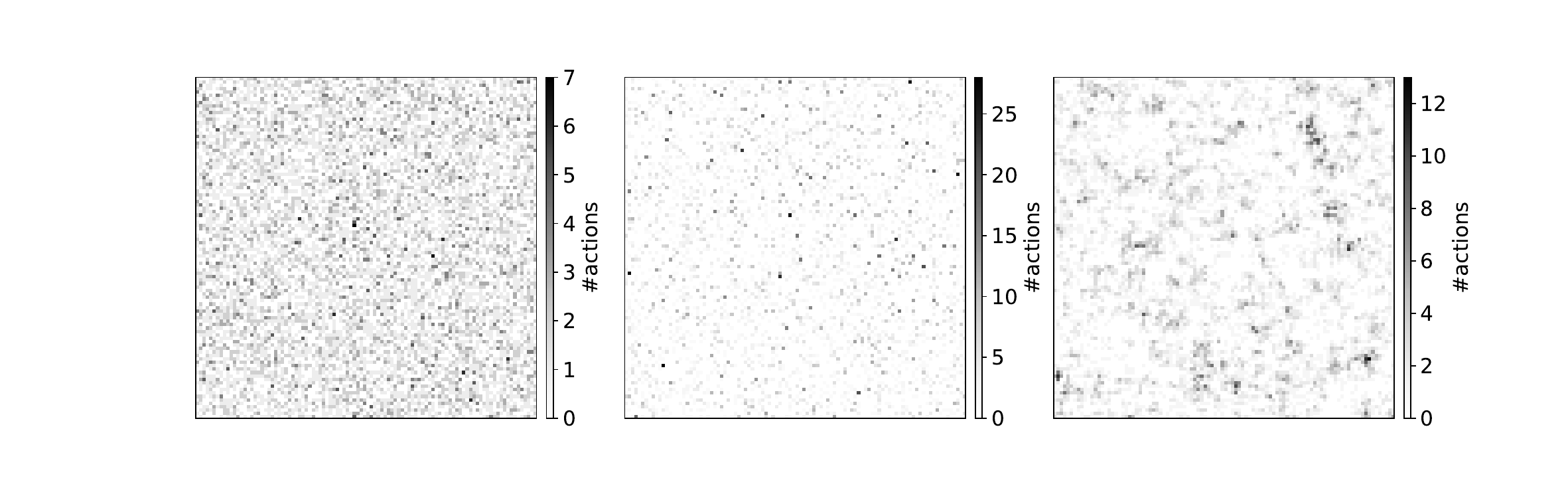}
	\caption{
        Distribution of the number of actions in three cases. 
        }
	\label{fig:action_distribution}
\end{figure}

We attributed this result to the non-uniformity of the number of actions. 
This idea is based on the fact that the distribution of the number of times was non-uniform, both totally and spatially, as shown in Fig.~\ref{fig:action_distribution}.
The following subsection will investigate this in detail. 

\subsection{Experiment II: Analyzing the effect of non-uniformity in the number of donations}
We considered that the differences among the above results are caused by the difference in non-uniformity in the number of donations.
To validate the idea, we initially prepared some indices of non-uniformity and subsequently examined the relationships between them and the evolution of cooperators. 

We defined three indices: 
1) standard deviations ($\sigma_{\rm d}$) of the number of donations; 
2) decay rates ($\gamma_1$) obtained by fitting the distribution of the number of donations to a power-law distribution; 
3) averaged similarities of the number of donations between neighbors, which we call ``donation correlation ($r_{\rm d}$)." 
The calculus of $\sigma_{\rm d}$ is as follows: 
\begin{align}
	\sigma_d &= \frac{1}{N-1} \sum^{N}_{i=1} (d_i - \overline{d})^2, 
\end{align}
where $d_i$ represents the number of donations of agent $i$. 

\begin{figure}[htbp]
\centering
\includegraphics[width=0.7\columnwidth]{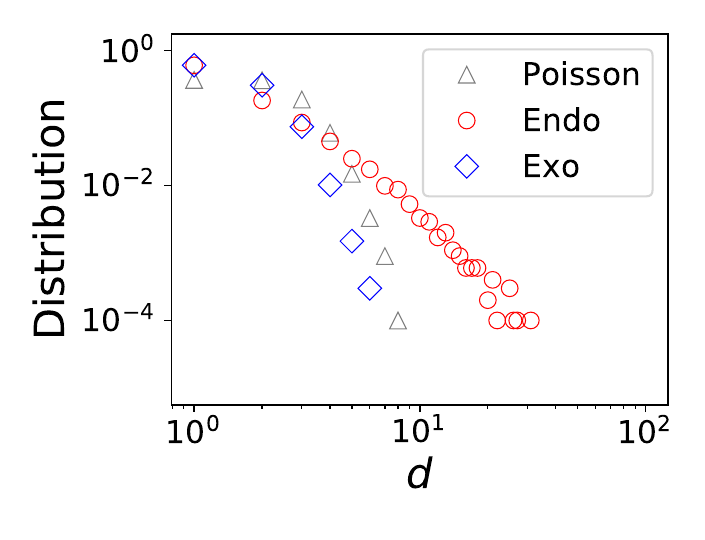}
\caption{Log-log plots of the distribution of donation actions for each of the Poisson (gray triangles), Endo (red circles), and Exo (blue diamonds) cases.}
\label{fig:distributions}
\end{figure}

The idea of $\gamma_1$ came from the observation of distributions of donations in the cases shown in Fig.~\ref{fig:distributions}. 
The gray triangles, the red circles, and the blue diamonds indicate the Poisson, Endo, and Exo cases, respectively. 
Both axes are logarithmic. 
This figure shows that the distribution of the Endo case seemed to follow a power-law distribution known to have a fat tail. 
Here, we formulate the power-law distribution as $D(d) = \gamma_0 d^{-\gamma_1}$, where $\gamma_0$ and $\gamma_1$ are coefficients.
Although the Poisson and Exo cases seem to follow exponential distributions, we can also regard them as a power-law distribution with a thinner tail.
Generally, a power-law distribution becomes the uniform distribution as its tail becomes fatter ({\it {\it i.e.}}, as $\gamma_1$ decreases). 
In contrast, it becomes the delta distribution as its tail becomes thinner (as $\gamma_1$ increases). 
In terms of the uniformity of the values produced, it is clear that the uniform distribution is the most non-uniform, and the delta function is entirely uniform.
Therefore, we decided to use $\gamma_1$ obtained by fitting distributions of $d_i$ to the power-law distribution as a parameter that reflects the non-uniformity. 

Finally, inspired by the degree correlation in network science, we computed the donation correlation $r_d$, which indicates the local similarity in the number of actions, using the following formula: 
\begin{align}
r_d =\left\{
	\begin{array}{ll}
		\frac{\sum_{(u,v)\in E} (d_u d_v -\overline{d}^2)}{|E| (\overline{d^2}-\overline{d}^2)}  &  (\overline{d^2}\ne\overline{d}^2), \\
		1  &  ({\rm otherwise}).
	\end{array}
\right.
\label{eq:relation}
\end{align}
In Eq.~\eqref{eq:relation}, $E$ is the set of all edges on the lattice, and $|E|$ indicates the total number of edges calculated by $kN/2$ using the number of neighbors $k$ and the total number of agents $N$. 
Moreover, $d_u$ is the agent $u$'s number of donation actions, $\overline{d}$ is the average of $d_u$, and $\overline{d^2}$ is the average of squared $d_u$. 
The values of $r_d$ range from -1 to 1; 
it indicates that the number of donations between neighbors becomes different as $r_d$ becomes closer to -1, while the number becomes similar as $r_d$ becomes closer to 1. 
Thus, the standard model shows us $r_d=1$. 
In contrast, in a particular case where individuals with $d=0$ and $d=1$ are distributed like a chessboard, $r_d$ takes on the value of -1.

\begin{figure}[htbp]
\centering
\includegraphics[width=\columnwidth]{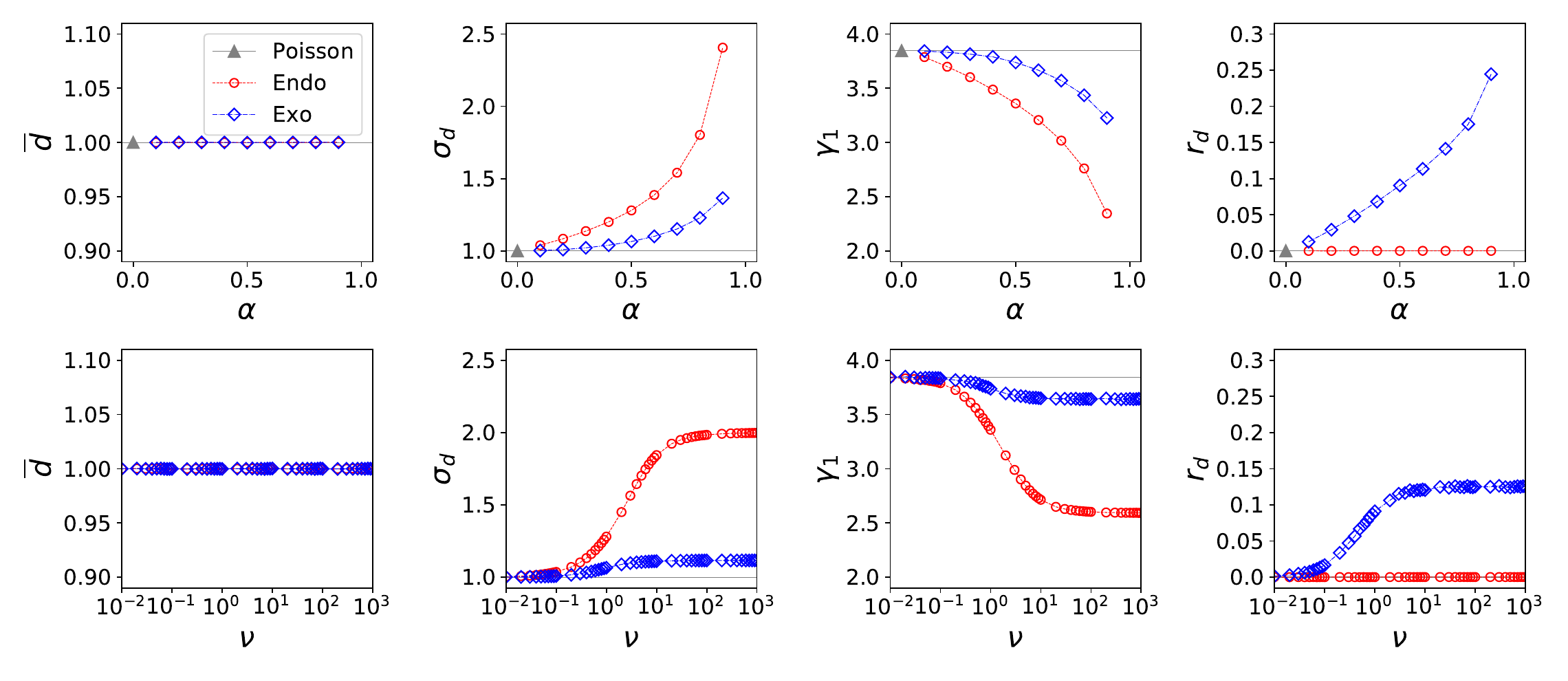}
\caption{The relationship between parameters related to behavior and indices of non-uniformity. The upper row takes $\alpha$ as the horizontal axis, while the lower row takes $\nu$ as the horizontal axis.}
\label{fig:non-uniformity}
\end{figure}

Before analyzing the relationship between these indices and the evolution of cooperation, 
we will clarify how the intensity of endogenous and exogenous cascades affect those indices. 
The upper (lower) row of Fig.~\ref{fig:non-uniformity} plots the changes in the non-uniformity indices as $\alpha$ ($\nu$) is changed for the Endo and Exo cases. 
The first column represents the average of donation actions in the population. 
In this experiment, we adjusted $\rho$ to keep the average value constant to focus solely on non-uniformity. 
We can confirm that this condition has been achieved. 
The second, third, and fourth columns indicate $\sigma_d$, $\gamma_1$, and $r_d$, respectively. 
Additionally, it is essential to note that the non-uniformity indices are based on $b=1.1$ (keeping in mind that the distribution of donation frequency is or is not independent of cooperation and thus unrelated to $b$). 
The gray line with a triangle marker illustrates the results of the Poisson case.

Observations from this figure can be summarized as follows:
1) In the first column, we can ensure that the averages of donation actions ($\overline{d}$) are always one; 
2) The standard deviations $\sigma_d$ (the second column) increased as $\alpha$ and $\nu$ increased, but the decay rates $\gamma_1$ (the third column) decreased; 
3) In addition, the Endo case changes more dramatically than the Exo. 
4) The fourth column, $r_d$, shows that only the Exo case increases though others lie on zero. 

From these observations, we can propose hypotheses to explain the order among the cases illustrated in Fig.~\ref{fig:Result1}: 
1) The results of the Endo case can be attributed to a significant global non-uniformity ($\sigma_d$ and $\gamma_1$) in the number of donations;
2) The outcomes in the Exo case are caused by the local similarity ($r_d$) in the number of donations.

\begin{figure}[htbp]
\centering
\includegraphics[width=\columnwidth]{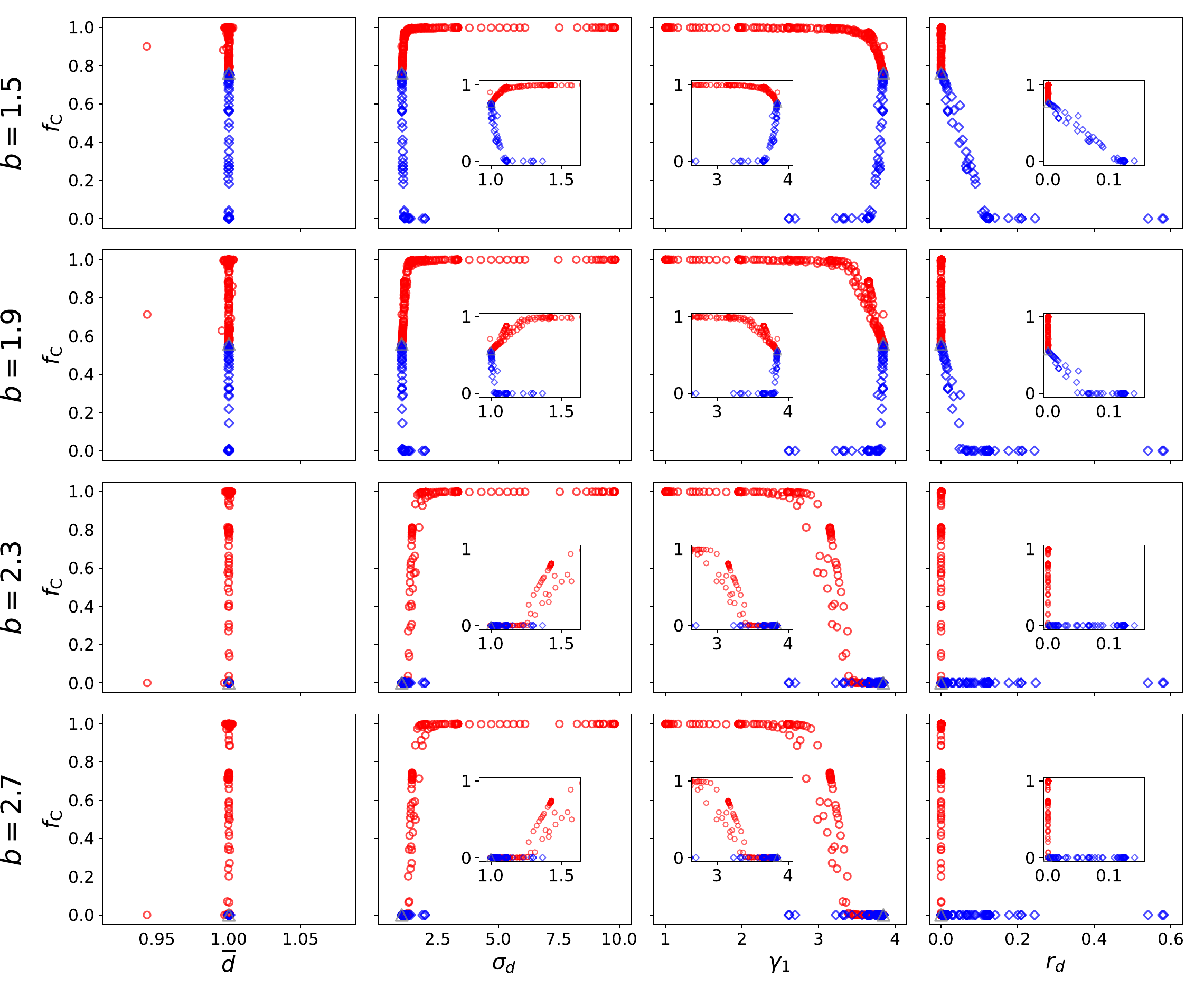}
	\caption{Relationships between three non-uniformity indicators (the standard deviations $\sigma_d$, the decay rate of power law $\gamma_1$, and the local similarity in the number of donations $r_d$) and the fraction of cooperators by the scatter plot in each case of $b=1.5$, $1.9$, $2.3$, and $2.7$. }
\label{fig:non-uniformity-scatter}
\end{figure}

Figure~\ref{fig:non-uniformity-scatter} shows the relationships between each non-uniformity indicator and the fraction of cooperators by the scatter plot in each case of $b=1.5$, $1.9$, $2.3$, and $2.7$. 
Each point represents the results obtained by varying $\alpha$ and $\nu$ while keeping $b$ fixed, with an average of 100 trials for each point.
The red points, blue points, and gray triangles represent the results of the Endo, Exo, and Poisson cases, respectively. 
The internal graph is an enlarged portion to confirm the linearity.
Let's focus on the Endogenous case first.
Because $r_d$ (local non-uniformity) is consistently zero and thus irrelevant, we can focus on the relationship between $\sigma_d$ and $\gamma_1$ (global non-uniformity) and the evolution of cooperation in the Endo case.
Examining the internal graphs, we can confirm partial positive linearity with $\sigma_d$ and partial negative linearity with $\gamma_1$.
Therefore, we can observe that more significant global non-uniformity has an enhancing effect on the evolution of cooperation.
Furthermore, by comparing each row, it is indicated that the impact on the evolution of cooperation diminishes as $b$ increases.
We also see that for $b > 2$, the condition $\sigma_d > 1.2$ ($\gamma_1 < 3.4$) was imposed to preserve the impact.
%
Conversely, in the subplot of $b = 1.5$ in the fourth column ($r_d$) shows linearity in the region of approximately $r_d < 0.13$, indicating that the cooperator survives. 
Similarly, the subplot for $b = 1.9$ shows the condition for $r_d < 0.04$. 
For $b > 2$, the condition of $r_d$ under which the cooperator survives in the Exo case disappears. 
Considering that Fig.~\ref{fig:non-uniformity} shows lowering $r_d$ (i.e., decreasing $\alpha$ in Exo) reduces global non-uniformity metrics like $\sigma_d$ or $\gamma_1$, we can infer that sufficiently low $r_d$ fails to meet the global non-uniformity conditions observed in the Endo case.
It can be considered that a sufficiently low $r_d$ no longer satisfies the condition for global non-uniformity observed in the Endo case.
Therefore, it is suggested that larger local non-uniformity has a negative effect on the evolution of cooperation and that this effect exceeds the positive effect of global non-uniformity.
In other words, it suggests that more significant local non-uniformity hurts the evolution of cooperation, and this effect outweighs the positive impact associated with the global one.

From the observations above, we can suggest that the degree of variability in standard deviation ($\sigma_d$), power-law distribution ($\gamma_1$), and donation correlation ($r_d$) is linearly related to the proportion of cooperators in the particular region. 
The comparison between the Exo and Endo cases reveals that low local non-uniformity is a more preferred factor in the evolution of cooperation. 
This finding is natural because local non-uniformity can affect agents' actions and decisions more than global ones, given the spatial constraints imposed on the agents' actions.
In addition, since the global non-uniformity results show that the more severe the non-uniformity is, the more cooperation is promoted, the relationship between non-uniformity and cooperation we find can be summarized as ``cooperation is encouraged when the number of donations differs among neighbors more often and when the differences are more varied."

Why does the relationship between non-uniformity and the evolution of cooperation appear? 
We consider it is because non-uniformity allows for cooperators with an extreme number of actions. 
Let us first consider cooperators with $d_i=0$ (hereafter called ${\rm C}_0$). 
Despite having $a_i={\rm C}$, ${\rm C}_0$ has never had the opportunity to act, so they do not change their or their neighbors' payoffs. 
Therefore, ${\rm C}_0$ is neutral in the evolution concerning non-cooperators; in other words, it can resist non-cooperators without forming clusters with other cooperators. 
Due to their numerous donations, these cooperators bear significant costs and are thus evolutionarily weak. 
Conversely, the more they contribute, the more likely their neighbor agents will survive. 
Therefore, the pair of ${\rm C}_0$ and ${\rm C}_n$ appears advantageous for cooperators.

Analyzing the frequency of this combination can help us understand the relationship between non-uniformity and the evolution of cooperation. 
Let us consider local non-uniformity and the tendency for ${\rm C}_0$-${\rm C}_n$ pairs to appear. 
When local non-uniformity is at its maximum ({\it i.e.}, 1), adjacent agents have an equal number of actions, so this pair cannot exist, and cooperation struggles to survive in the standard model (dashed line in Fig.~\ref{fig:Result1}). 
When local non-uniformity is high, the number of actions of adjacent agents are similar, making the appearance of such pairs rare.
Conversely, when local non-uniformity is low, adjacent agents have a different number of actions, increasing the likelihood of the pair's occurrence. 
Moreover, local non-uniformity exhibits a threshold for the evolution of cooperation (as seen in the 4th column of Fig.~\ref{fig:non-uniformity-scatter}). 
We believe this threshold exists because if an insufficient number of pairs are created to counteract the effect of non-cooperators, cooperators go extinct.
As global non-uniformity increases, cooperators who have made more contributions emerge, thus increasing the probability of the emergence of more strong ${\rm C}_0$-${\rm C}_n$ pairs. 

Hence, we argue that the probability of the emergence of ${\rm C}_0$-${\rm C}_n$ pairs in the state of non-uniformity can explain the relationship between non-uniformity and the evolution of cooperation.

\section{Discussion}
To gain a deeper understanding of the mechanisms behind human behavioral change (cultural evolution), we constructed a multi-agent model where agents act according to timings determined by the Hawkes process and investigated the impact of cascading actions on the evolution of cooperation. 
In this study, we employed two mechanisms for cascading actions: self-excitation (Endo), where an individual's past actions influence their current actions, and external excitation (Exo), where the past actions of others affect an individual's current actions.
Additionally, we included the other two comparative cases: one where agents act randomly according to the Poisson process (Poisson), and the other representing biological evolution (standard), where actions are performed only once.

Simulation results initially demonstrated the ranking of these four cases, revealing that Endo most effectively promotes cooperation. 
We attributed this result to the differing distributions of the number of actions associated with each behavioral rule.
Subsequently, we defined a measure of non-uniformity and examined its relationship with the evolution of cooperation. 
Our findings indicated that when local non-uniformity (non-uniformity among neighbors) is low, higher global non-uniformity promotes cooperation more effectively.
The likely reason for this condition is that it facilitates the formation of pairs consisting of cooperators who have never acted and those who have acted extensively. 
These results suggest that random elements causing non-uniformity and cooperation by a few highly active individuals effectively maintain a cooperative society. 
Conversely, the tendency to follow others' actions does not help to escape the temptation of free-riding.

We confirm that local and global non-uniformity in the frequency of actions is a crucial determinant in the evolution of cooperation. This leads us to the following assumptions.
First, actions triggered by others are disadvantageous. 
However, in the real world, people are often influenced by others to act, so this assumption contradicts real-world phenomena. 
Therefore, it is necessary to examine more deeply whether there is a realistic setting in which cooperation can evolve, which was unintentionally ignored in our model, even in Exo cases. 
Alternatively, it is essential to explore whether there is some advantage of Exo cases from another perspective that remains characteristic of human behavior.
%
%
Second, the result indicating that greater local non-uniformity favors the evolution of cooperation raises the question of whether it is possible to create a more locally non-uniform situation using the Hawkes process model. 
In other words, the smallest value of local non-uniformity was zero for Endo, Poisson, and Exo. 
However, is it possible to create a situation where the value is lower than zero, or is such a situation nonexistent in reality? 
In this context, we consider an extended model where the actions of others suppress the occurrence of an individual's action.

On the other hand, it is important to note that this study only examines limited situations. 
One significant limitation is the fixed spatial structure of the square lattice. 
Since spatial structure significantly influences the results, examining the relationship with spatial structure in more detail is essential. 
However, a technical problem arises when applying the Hawkes process directly to other well-known networks, such as scale-free and small-world networks, due to potential divergence issues. 
If we can solve this problem, we can test various networks.
%
Furthermore, while we used identical excitation parameters for all agents in this study, allowing these parameters to evolve could reveal the optimal excitation levels.

\section*{Acknowledgment}
This study was partly supported by JSPS KAKENHI, Grant No. JP23K22982 (G.I.).

\section*{Declaration of generative AI and AI-assisted technologies in the writing process}
While preparing this work, the authors used ChatGPT-4o to improve the readability and language of the manuscript. 
After using this tool, the authors reviewed and edited the content as needed. The authors take full responsibility for the content of the published article.

\appendix
\setcounter{equation}{0}
\renewcommand{\theequation}{\Alph{section}\arabic{equation}}

\section{\label{app:expected_intensity}Expected intensity}
First, consider the expected intensity when there is a single event generator. 
In the Hawkes process, the branching ratio is defined as $\int_0^{\infty} g(\tau) {\rm d}\tau$. 
This indicator represents how many events an event occurrence causes on average. 
When the Hawkes process follows Eq.~\eqref{eq:ExpHawkes}, its branching ratio can be calculated as follows:
\begin{align}
	\gamma &= \int_0^{\infty} \alpha \nu \exp(-\beta\nu\tau) {\rm d}\tau \nonumber \\
	&= \frac{\alpha}{\beta}. 
	\label{eq:branching_ratio_single}
\end{align}
Using this, we can obtain the expected intensity $\left<\lambda\right>$, {\it i.e.}, the average number of events per unit time: 
\begin{align}
	\left<\lambda\right> &= \sum_{k=0}^{\infty} \gamma^k = \frac{1}{1-\gamma}, 
	\label{eq:expected_intensity_single}
\end{align}
where $\gamma<1$ must be satisfied. 

Second, we deal with the case where multiple event generators are considered, which is also used in our multi-agent model. 
Hereafter, we regard event generators as individuals according to the model. 
Using Eq.~\eqref{eq:lambda_origin}, we calculate the effect of an excitation of individual $i$ on $j$, {\it i.e.}, the branching ratio $\gamma_{i,j}$ from $i$ to $j$. 

In the case of $i=j$, we obtain
\begin{align}
	\gamma_{ij} &= \int_0^{\infty} \alpha_{j;n} \nu_j \exp(-\beta_{j;n} \nu_j \tau) {\rm d}\tau \nonumber \\
	&= \frac{\alpha_{j;n}}{\beta_{j;n}}. 
\end{align}
Then, in the case of $j\in N_i$, we obtain
\begin{align}
	\gamma_{ij} &= \int_0^{\infty} \alpha_{j;x} \nu_j \exp(-\beta_{j;x} \nu_j \tau) {\rm d}\tau \nonumber \\
	&= \frac{\alpha_{j;x}}{\beta_{j;x}}. 
\end{align}
Otherwise, $\gamma_{ij}=0$. 
%

The $N\times N$ matrix of $\gamma_{ij}$ is defined as $\Gamma$. 
It can be regarded as an $N\times N$ lattice adjacency matrix with the weights of the branching ratios, or more specifically, the amount of influence of individual $i$ on individual $j$. 
Using this, the $N\times1$ column vector of expected values for each individual $\left<\bm\lambda\right>$ composed of $\left<\lambda_i\right>$ is obtained by
\begin{align}
	\left<\bm\lambda\right> &= (I-\Gamma)^{-1} \bm\rho, 
	\label{eq:expected_intensity_inverse}
\end{align}
where $I$ is the $N\times N$ identity matrix and $\bm\rho$ is the column vector of $\rho_i$. 
%

This time, we use Gershgorin's theorem to estimate the sufficient conditions under which the spectral radius of $\Gamma$ is smaller than 1. 
Using the standard parameters $\alpha_n,\ \alpha_x\in[0,1)$ and $\beta>0$ and then simplifying $\alpha_{i;n}=\alpha_n$, $\alpha_{i;x}=\alpha_x$, $\beta_{i;X}=\beta\ (X\in\{n,x\})$, we obtain $D(\frac{\alpha_{n}}{\beta}, k\frac{\alpha_{x}}{\beta})$, which is the only Gershgorin disk of $\Gamma$. 
Therefore, from $\frac{\alpha_n}{\beta}>0$, we know that the eigenvalues of $\Gamma$ are at most $\frac{\alpha_n+k\alpha_x}{\beta}$ and at least $\frac{\alpha_n-k\alpha_x}{\beta}$.

Define $\beta\equiv o_n+k o_x$, where $o_n\in\{0,1\}$ and takes the value 1 when $\alpha_n>0$ and 0 when otherwise. 
$o_x$ has the same meaning for $\alpha_x$.
Then, we obtain
\begin{align*}
	\frac{\alpha_n+k\alpha_x}{o_n+ko_x}<1,\ \frac{\alpha_n-k \alpha_x}{o_n+ko_x}>-1. 
\end{align*}
Thus, the spectral radius of $\Gamma$ is smaller than 1 when using these definitions. 

Let us transform Eq.~\eqref{eq:expected_intensity_inverse}. 
Suppose we use $\rho_i$ with the same value for all individuals. 
That is, $\rho_i=\rho$ using the common parameter $\rho$.
\begin{align}
	\left<\bm\lambda\right> &= (I-\Gamma)^{-1} \bm\rho \nonumber \\
	&= (I-\Gamma)^{-1} \rho \bm1 \nonumber \\
	\therefore \rho \bm1 &= (I-\Gamma) \left<\bm\lambda\right> \nonumber \\
\end{align}
%
%

Here, if we assume that $\left<\lambda_i\right>$ has the same value for all individuals ($\left<\lambda\right>$), 
\begin{align}
\rho\bm1 &= (I-\Gamma)\left<\lambda\right>\bm1 \nonumber \\
\therefore \rho &= \left<\lambda\right> \left(1-\frac{\alpha_n}{\beta}-k\frac{\alpha_x}{\beta}\right) \nonumber \\
&= \left<\lambda\right> \left(1-\frac{\alpha_n+k\alpha_x}{\beta}\right). \label{eq:adjust_rho}
\end{align}

\bibliographystyle{unsrt}

\end{document}